\documentclass[sigconf,nonacm]{acmart}
\AtBeginDocument{%
  }

\renewcommand\footnotetextcopyrightpermission[1]{}
\setcopyright{none}

\usepackage{listings}
\usepackage{xcolor}
\definecolor{risk}{HTML}{BD0A0A}      
\definecolor{balanced}{HTML}{4059ad}  
\definecolor{loss}{HTML}{61b96b}      
\definecolor{erratic}{HTML}{595959}   

\begin{document}

\title{Biased Error Attribution in Multi-Agent Human–AI Systems Under Delayed Feedback}

\author{Teerthaa Parakh}
\affiliation{%
  \institution{Georgia Institute of Technology}
  \city{Atlanta}
  \state{GA}
  \country{USA}
}

\author{Karen M. Feigh}
\affiliation{%
  \institution{Georgia Institute of Technology}
  \city{Atlanta}
  \state{GA}
  \country{USA}}

\renewcommand{\shortauthors}{Parakh et al.}

\begin{abstract}

Human decision-making is strongly influenced by cognitive biases, particularly under conditions of uncertainty and risk. While prior work has examined bias in single-step decisions with immediate outcomes and in human interaction with a single autonomous agent, comparatively little attention has been paid to decision-making under delayed outcomes involving multiple AI agents, where decisions at each step affect subsequent states.

In this work, we study how delayed outcomes shape decision-making and responsibility attribution in a multi-agent human–AI task. Using a controlled game-based experiment, we analyze how participants adjust their behavior following positive and negative outcomes. We observe asymmetric responses to gains and losses, with stronger corrective adjustments after negative outcomes. Importantly, participants often fail to correctly identify the actions that caused failure and misattribute responsibility across AI agents, leading to systematic revisions of decisions that are weakly related to the underlying causes of poor performance.

We refer to this phenomenon as a form of attribution bias, manifested as biased error attribution under delayed feedback. Our findings highlight how cognitive biases can be amplified in human–AI systems with delayed outcomes and multiple autonomous agents, underscoring the need for decision-support systems that better support causal understanding and learning over time.

\end{abstract}

\ccsdesc[500]{Human-centered computing~Empirical studies in HCI}
\ccsdesc[300]{Computing methodologies~Artificial intelligence}
\ccsdesc[100]{Computing methodologies~Modeling and simulation}
\ccsdesc[500]{Computing methodologies~Multi-agent systems}

\keywords{Cognitive bias, human decision making, human–AI interaction, empirical study, multi-agent systems, reinforcement learning}


\maketitle

\section{Introduction}
Human decision-making is influenced by cognitive biases, defined as systematic deviations from fully analytic judgment, particularly under conditions of uncertainty and risk \cite{tversky1974judgment,tversky1992advances}. These biases emerge from the reliance on heuristics during decision-making. Heuristics are often adopted in response to complex environments, time pressure, or information overload that necessitates selective attention \cite{bias_search}. Although heuristics can reduce cognitive effort, they frequently introduce systematic judgment errors. Data from the European Helicopter Safety Team show that the majority of fatal crashes are attributable to decision errors rather than perceptual or execution errors (European Union Aviation Safety Agency, 2012). Therefore, cognitive biases can have serious consequences in high-stakes domains such as aviation, healthcare, and medical decision-making.

The presence of automated or AI agents can further amplify existing biases, such as anchoring and automation bias \cite{Endsley2017HFES,Rosbach2025TwoWrongs,10.1145/3290605.3300831}, and may also introduce new forms of bias during decision-making. For example, when users are presented with a correct AI suggestion first, they tend to rely more heavily on the AI and exhibit stronger automation bias compared to users who first see an incorrect suggestion, demonstrating ordering effects \cite{10.1145/3397481.3450639}.

Therefore, cognitive biases may originate from how information is presented to users, what information is shown, when it appears (ordering effects), and how it relates to users’ prior beliefs (confirmation bias), and these factors can strongly shape users’ reliance on AI \cite{medical_cognitive_bias,medical_search,bias_search}. The presence of these biases can lead to either over-reliance on AI or under-reliance on AI. Both outcomes are problematic: over-reliance occurs when users continue to trust the AI even when it is incorrect, while under-reliance occurs when users fail to use AI to its full potential, which may reduce overall team performance \cite{passi2022overreliance}. One of the main causes of over-reliance is a lack of domain knowledge on the part of the user. The issue of over-reliance becomes especially important in high-stakes domains where safety is critical. Approaches such as explainable AI (XAI) \cite{10.1145/3290605.3300831}, nudging, and boosting \cite{caraban2019ways,nudging_boosting,web_single_step} have been proposed to mitigate cognitive biases in such scenarios.

However, most prior work \cite{10.1145/3287560.3287563, 10.1145/3397481.3450639, 10.1145/3290605.3300831, 10.1145/3503252.3534362, web_single_step} focuses on single-step decision-making scenarios or interaction between a human and a single AI agent, with comparatively less attention given to cascaded decision-making scenarios or interaction with multiple AI agents. For example, \cite{10.1145/3397481.3450639} studied a task in the cooking domain where participants were asked whether a set of kitchen policies was being followed. \cite{medical_cognitive_bias,medical_search,hypothesis_medical,10.1145/3290605.3300831} examined tasks in clinical or medical domains, where \cite{10.1145/3290605.3300831} studied the use of XAI for intensive-care phenotyping, and \cite{10.1145/3503252.3534362, web_single_step} examined tasks in the web search domain.

In this work, we examine how cognitive biases manifest in cascaded decision-making involving a human and multiple autonomous agents. Unlike single-step decisions, cascaded decision-making involves delayed feedback or delayed outcomes and decisions made at one step influence the subsequent state. Interaction with multiple autonomous agents further increases complexity, since tasks become interdependent and overall success depends on coordinated performance. Such situations are common in command-and-control settings, including military mission planning, air traffic control, and emergency response management. Given the frequent occurrence of such scenarios, it is important to study how cognitive biases appear in these settings and how they influence decision-making.  

To study this, we developed a game environment that incorporates cascaded decision-making and involves humans interacting with multiple AI agents. We analyze participants’ gameplay to examine how delayed feedback together with cascaded decisions influences decision-making in multi-agent human–AI tasks. We find that when outcomes are observed only after a sequence of actions, participants often fail to correctly identify which decisions led to suboptimal results and fail to accurately assign responsibility to the specific decisions that produced the outcome. This difficulty parallels the credit assignment problem under sparse and delayed rewards in reinforcement learning \cite{pignatelli2024surveytemporalcreditassignment,credit_assignment_sutton}, where delayed feedback obscures the causal contribution of individual actions.

The difficulty humans face in assigning responsibility under delayed feedback is further compounded by the presence of multiple AI agents providing guidance during decision-making. Our behavioral analyses indicate that participants struggle to determine which AI agent’s recommendations should be adjusted following failure. Instead, they tend to attribute negative outcomes to one of the AI agents. We interpret this pattern as evidence of attribution bias, expressed as biased error attribution under delayed feedback in multi-agent human–AI decision-making.

\section{Background}

Research in domains such as search and healthcare indicates that the way information is presented can strongly shape user judgments, especially when users lack domain expertise \cite{medical_cognitive_bias,medical_search,bias_search}. In such cases, users often focus on salient or alarming information, seek evidence that confirms their prior beliefs, and misinterpret technical details, which can lead to confirmation bias and incorrect conclusions even when accurate information is available \cite{hypothesis_medical}.

Responsibility attribution in human–AI interaction is also affected by cognitive biases. Prior work reports that the way blame or credit is assigned between humans and AI depends on the outcome valence. For example, \cite{10.1145/3706598.3713126} found that positive outcomes are more likely to be associated with shared responsibility, whereas negative outcomes tend to lead users to attribute responsibility to a single entity, though not consistently to either the human or the AI.

To address such biases, several mitigation approaches have been proposed. Techniques such as nudging, boosting, and explainable AI have been suggested as ways to encourage more reflective decision-making \cite{caraban2019ways,nudging_boosting}. However, later work indicates that methods intended to improve human–AI interaction can also introduce unintended effects. In particular, explanations and interpretability tools may amplify anchoring bias and reinforce existing beliefs, leading users to overestimate or underestimate system capabilities \cite{anchoring_explanation,confirmation_explanation,reliance_explanation,workshop_biases}. Other studies report that users may over-trust explanations, misuse visualizations, or fail to develop accurate mental models of the system, resulting in inappropriate reliance rather than improved decision-making \cite{10.1145/3397481.3450639,10.1145/3313831.3376219}.

Studying cognitive bias in human–AI interaction is complicated by differences across users. Prior work notes that humans vary in cognitive limitations, prior experience, and expertise, and that a single individual may exhibit multiple biases simultaneously, making such effects difficult to isolate and mitigate \cite{workshop_biases}. As a result, mitigation tools such as explainable AI (XAI) often show limited effectiveness. In addition, although personalized and adaptive systems aim to better align with individual users, such personalization may further intensify existing biases. For example, systems that closely match user preferences, such as recommender systems or large language models, can reinforce confirmation bias or create echo-chamber effects by repeatedly presenting information consistent with prior beliefs \cite{survey_bias_debias,survey_echo,workshop_biases}.

Cognitive biases in human–AI interaction have been examined across several domains. Existing work primarily focuses on information interaction, recommender systems, visualization, behavior change, and usability, while comparatively less research considers Computer-Supported Cooperative Work scenarios, where interactions involve multiple agents and shared decision-making, and where decisions often unfold over multiple steps rather than a single interaction \cite{10.1145/3706598.3713450}.

When humans interact with multiple agents, system outcomes emerge from coordination among interdependent agents rather than from a single source. Prior research on multi-agent systems suggests that outcomes depend on the interaction among agents rather than on a single decision-maker \cite{prada2024sustainable}. In heterogeneous multi-agent systems, users must interpret not only individual outputs but also how agents influence one another, increasing the complexity of reasoning about responsibility and performance \cite{openai_agents_python_2025}. Such settings create conditions in which biases may affect how success or failure is attributed across agents.

Dynamic decision-making further increases this complexity. In dynamic environments, actions influence future states and feedback is often delayed, making cause–effect relationships difficult to infer \cite{martin2004actrbeer}. Although performance may improve through repeated interaction and strategy adaptation, decision-making remains learning-based and experience-driven \cite{martin2004actrbeer,prebot2023cyberdefense}. Building on these findings, we analyze consecutive games to examine how prior outcomes shape subsequent decisions multi-agent AI settings.

\section{Method}
We conducted an empirical assessment of human gameplay in a naval battle strategy game that satisfies two key requirements: (1) it involves multiple agents, and (2) it requires cascaded decision-making. To keep the problem tractable, we focused on a setting with two agents interacting with a human participant. In this scenario, the AI agents serve as offensive and defensive advisors to participants, who take on the role of commander. The game was developed using the GameTeq software \cite{metateq2025gameteq}. In the following sections, we describe the game setup in detail.

\subsection{Game Setup}
\begin{figure*}[h]
    \centering
    \includegraphics[width=0.6\linewidth]{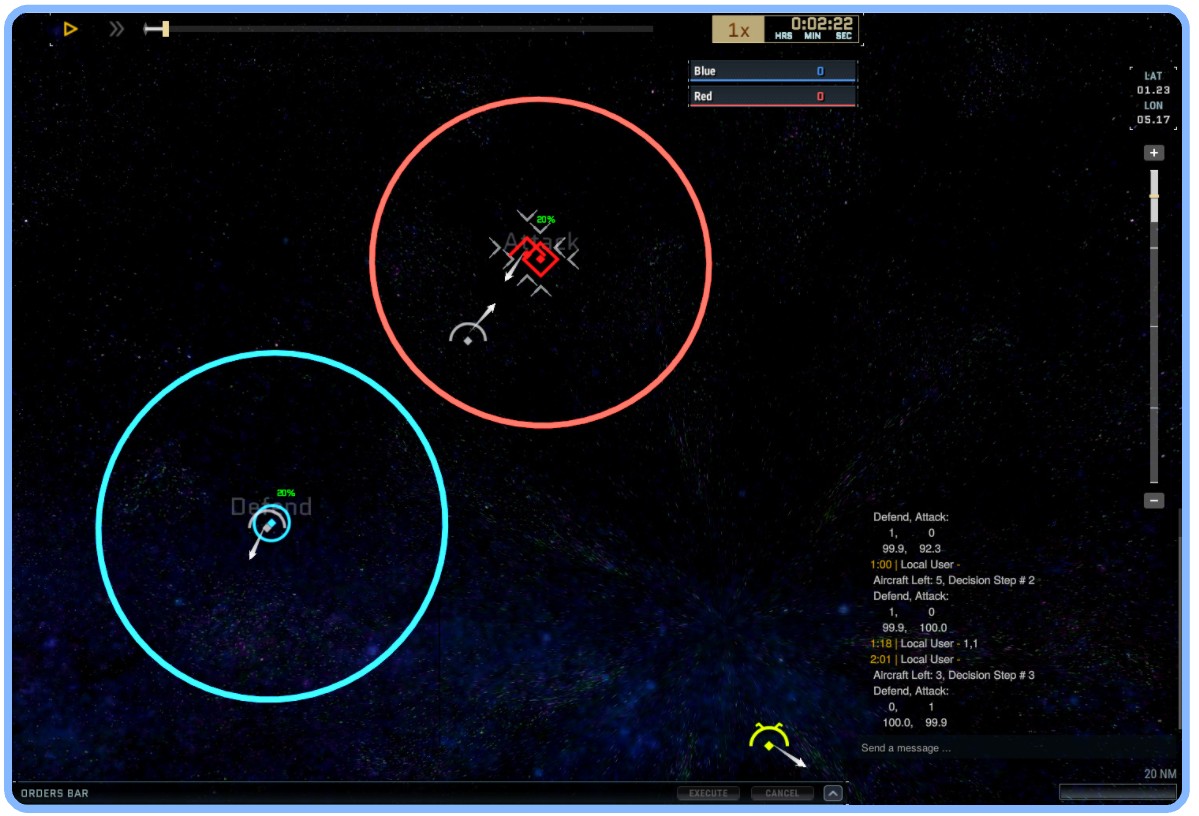}
    \caption{The game environment consists of two regions: the blue ally region and red adversary region. Each region contains a carrier at its center that forces must protect while attempting to destroy the opponent's carrier by launching aircraft. Enemies can enter the ally region at any time, requiring strategic resource allocation for both offensive and defensive operations. The ally side has a total of 5 aircraft available and must decide how many to deploy in each region across 7 time steps. The interface includes a clock and chat box in the bottom right where AI suggestions appear and participants enter their decisions, while scores for both forces are displayed in the top right corner.}
    \Description{game window screenshot}
    \label{fig:game}
\end{figure*}

The game is a resource allocation task with a limited number of resources and requires strategic decision making under time pressure. It models an offensive–defensive scenario involving two carrier strike groups: an \textcolor{blue}{ally group} and an \textcolor{red}{adversary group}. Each group consists of an aircraft carrier and multiple aircraft deployed at sea, as shown in Figure~\ref{fig:game}.Each side has five aircraft available.

The objective of the ally group is twofold: to defend the blue region from incoming adversary aircraft and to attack the adversary carrier located in the red region by deploying aircraft offensively. adversary aircraft follow a handcrafted policy, whereas the ally forces are guided by policies learned through reinforcement learning. Specifically, the ally side includes two AI agents serving as advisors: \textcolor{blue}{DefenseAI} and \textcolor{red}{OffenseAI}. \textcolor{blue}{DefenseAI} recommends how many aircraft should be deployed to defend the \textcolor{blue}{blue region}, while \textcolor{red}{OffenseAI} recommends how many aircraft should be allocated to the \textcolor{red}{red region}. 

Both AI agents are implemented as actor–critic networks trained using Proximal Policy Optimization (PPO) under a Centralized Training, Decentralized Execution (CTDE) framework \cite{amato2025initialintroductioncooperativemultiagent, multiagent_PPO}. Each agent’s actor network has partial observability and only observes information from its respective region (defense or offense). In contrast, the critic networks have access to the full game state.

\begin{figure}[h]
    \centering
    \includegraphics[width=\columnwidth]{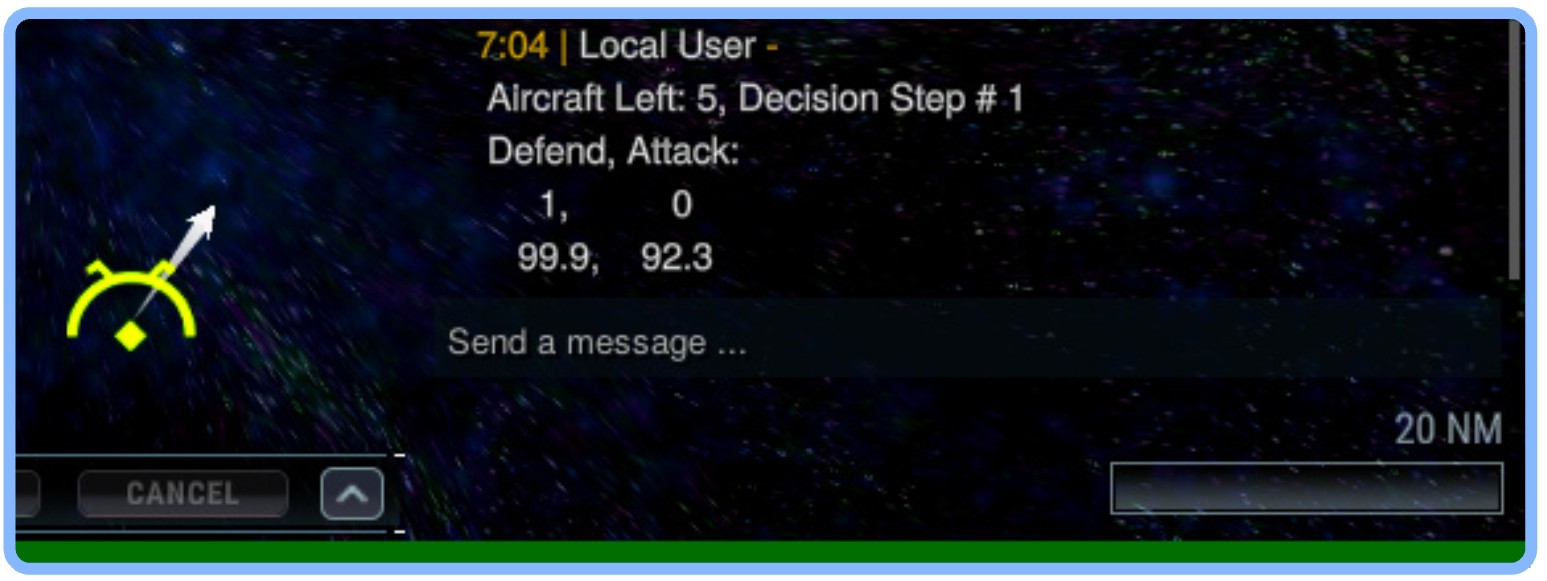}
    \caption{Chat interface displaying AI agent suggestions, remaining aircraft count, and current decision step in the game.}
    \Description{Chat interface displaying AI agent suggestions, remaining aircraft count, and current decision step in the game.}
    \label{fig:chat}
\end{figure}

The game proceeds in discrete time steps, with participants making decisions once every minute. Each game lasts for a maximum of seven minutes or until the health of either carrier is reduced to zero. At each decision step, participants receive recommendations from both AI agents via chat messages (Figure~\ref{fig:chat}), indicating the suggested number of aircraft to deploy in each region. This design restricts participants to high-level resource allocation decisions, while low-level actions—such as aircraft movement and adversary targeting—are either hard-coded or handled automatically by AI agents. Participants therefore act as commanders whose task is to allocate the five available aircraft across the two regions over the course of the seven minutes.

When making decisions, participants consider several factors, including adversary positions and health, the status and location of their own aircraft, and the recommendations provided by the AI advisors. They then enter their final allocation decision through the chat interface. Each decision must be made within 30 seconds while the game continues to run, introducing additional time pressure.

The game is intentionally asymmetric. For the ally side, each aircraft remains active for a maximum of two minutes after launch, unless its health reaches zero earlier. Once deactivated, the aircraft cannot be reactivated, but newly launched aircraft always start with full health. In contrast, adversary aircraft are launched with reduced health but remain active until their health reaches zero, with no time limit. This asymmetry encourages participants to deploy their aircraft gradually rather than launching all available aircraft at once.

Game difficulty is controlled by varying adversary strength, which includes the number of adversary aircraft and their health around the ally carrier at any given time. Participants can observe all adversary aircraft and their health in both the ally and adversary regions, allowing them to assess the current state of the game. The interface also displays a leaderboard showing the scores of both the ally and adversary forces. Scores are calculated based on the amount of damage inflicted on the opposing carrier. A higher score difference indicates better performance, as participants aim to maximize damage to the adversary carrier while minimizing damage to their own.

At the end of each game, a dialog box displays the final score. If the score is less than or equal to zero, it is displayed as zero; otherwise, the positive score is shown as achieved.

\subsection{Experiment Design}

We employed a mixed experimental design with three independent variables: (1) confidence score display (Confidence vs. No-Confidence), (2) game difficulty order, and (3) AI competency, which was either good AI (fully trained) or bad AI (poorly trained).

The first independent variable was the display of AI confidence scores. Participants were assigned to either a \textit{Confidence} or \textit{No-Confidence} condition, so this manipulation was between-subjects. In the Confidence condition, each AI agent (offense and defense) displayed an individual confidence score alongside its recommendation at every decision step. The confidence score was computed as the log probability of the action selected by the reinforcement learning policy and was intended to provide transparency into the agent’s decision-making. In the No-Confidence condition, participants received the same AI recommendations without any confidence information.

In second independent variable was the order of game difficulty levels to explore whether the timing of high-difficulty scenarios influenced participant behaviorThis manipulation was also between-subjects. In the initial version of the study, participants experienced the sequence (A) \textit{medium–\allowbreak low–\allowbreak high–\allowbreak medium–\allowbreak low–\allowbreak high}. In a later version, the order was changed to (B) \textit{medium–\allowbreak high–\allowbreak low–\allowbreak medium–\allowbreak low–\allowbreak high}, such that the relative order of the low-\allowbreak and high-\allowbreak difficulty games in the first half of the session was reversed. This manipulation was designed to examine whether encountering high-difficulty scenarios earlier versus later in the session influenced decision-making behavior in subsequent games and whether such effects persisted over time.

The third independent variable was AI competency, and it was within-subjects. For the medium- and high-difficulty games, we used a fully trained AI, and for the low-difficulty games, we used a poorly trained AI. The AI agents were trained in an environment mimicking the dynamics and stochasticity of the software and were trained without a human in the loop. When deployed in the interactive setting, the AI could behave suboptimally due to the increased stochasticity of the software and the presence of a human in the loop.

Table~\ref{tab:participant_num} summarizes the distribution of participants across the between-subject conditions.
\begin{table}[h]
  \centering
  \caption{Number of participants per experimental condition. Order A corresponds to the \textit{medium-low-high-medium-low-high} sequence, 
and Order B corresponds to the \textit{medium-high-low-medium-low-high} sequence.}
  \label{tab:participant_num}
  \begin{tabular}{lcc}
    \toprule
    \textbf{Condition} & \textbf{Order A} & \textbf{Order B} \\
    \midrule
    Confidence       & 8 & 5 \\
    No-Confidence    & 8 & 5 \\
    \bottomrule
  \end{tabular}
\end{table}

\subsection{Participants}

Participants were recruited for an in-person experiment and were required to be at least 18 years of age, fluent in English, and not color blind. All participants identified as video game players. A total of 30 participants took part in the study. Data from four participants were excluded from the analysis; three were excluded because they completed a version of the experiment in which all six game scenarios were identical, resulting in data that were not comparable to the rest of the sample. Data from one additional participant was excluded due to incorrect interaction with the interface that prevented meaningful task completion. All reported data and analyses are based on the remaining 26 participants.

The final sample included 13 participants who identified as male, 12 who identified as female, and 1 who preferred not to disclose their gender. Participants’ ages ranged from 18 to 39 years, with 11 participants in the 18-24 age group and 15 participants in the 25-39 age group.

Participants were compensated for their time with a payment ranging from \$15 to \$25, depending on their performance across the experimental games. The study protocol was approved by the institutional review board, and all participants provided informed consent prior to participation.

\begin{figure*}[h]
  \centering
  \includegraphics[width=0.85\linewidth]{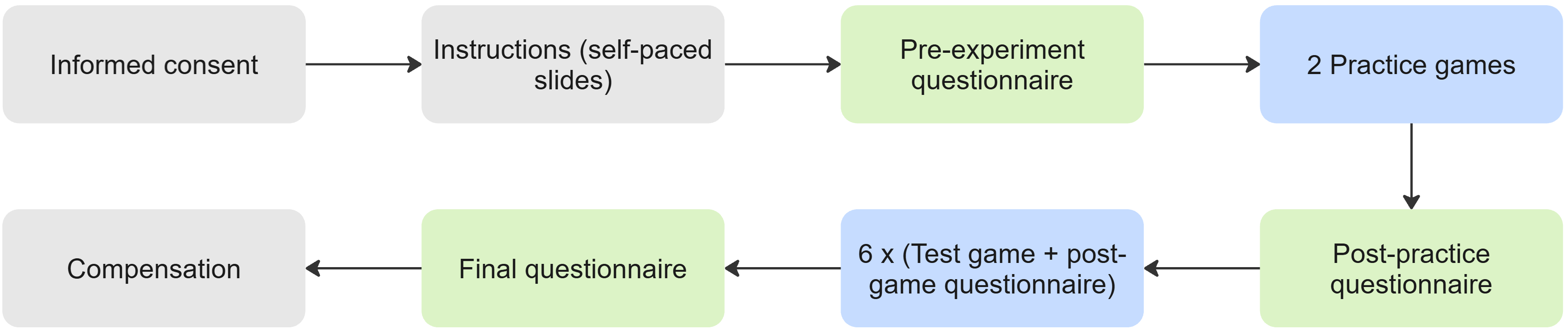}
  \caption{Overview of the experimental procedure.}
  \Description{A flowchart illustrating the experimental procedure. The chart shows sequence of study phases, including consent, instructions, questionnaires, practice games, six repeated test game–questionnaire cycles, a final questionnaire, and compensation, connected by arrows indicating the order of occurrence.}
  \label{fig:experiment_flow}
\end{figure*}

\subsection{Procedure}
After providing informed consent, participants were briefed on the study, the game mechanics, their task, and the AI advisors via self-paced instructional slides. Participants were assigned to one of two interface conditions (Confidence or No-Confidence) and one of two game difficulty orders. Participants then completed a pre-experiment questionnaire, played two practice games for familiarization, followed by a brief post-practice questionnaire, before beginning the main study. Next, participants played six test games, during which they received recommendations from AI agents at each decision point; confidence information was shown only in the Confidence condition. Participants completed a brief questionnaire after each game and a final questionnaire after completing all six games. Data from only the six test games were included. Sessions lasted approximately 60-75 minutes. Figure~\ref{fig:experiment_flow} summarizes the study procedure.

\subsection{Data Collection}
\label{sec:data_collection}
In our experiment, we collected data corresponding to both objective and subjective dependent variables. 
\paragraph{Objective Measures} We recorded the AI’s suggested actions, the actions taken by participants, the time taken by participants to make each decision, and the final game score at the end of each game. 
\paragraph{Subjective Measures} Subjective data were collected through questionnaires administered at multiple stages of the experiment. 

The pre-experiment questionnaire collected demographic information such as age range and gender, as well as background information including prior experience with AI systems and experience with video games. 

After each game, participants completed a questionnaire that included, among other items, ratings of how much they would trust each AI advisor using a 5 point Likert scale. Participants were also asked to indicate situations in which they believed each AI provided incorrect suggestions. These questions were designed to elicit participants’ mental models of the AI advisors. For the confidence condition, additional questions were included, such as indicating the level of confidence at which participants would be willing to accept an AI’s suggestion without independently evaluating the decision.

The final questionnaire included, among other items, participants’ overall level of trust in each AI advisor and their rating of how mentally demanding they found the decision-making process.

All questionnaires administered during the different phases of the study are provided in Appendix~\ref{app:questionnaires}.

\medskip
\noindent
Although the study included both independent and dependent variables, they are listed here for completeness. We collated all data from participants across all experimental groups. For the analysis of human behavior, we used the actions taken by participants and the game end score as the main outcome measures.

Our analyses focus on how strategy changes depend on prior game outcomes. The strategy definitions are described in Section~\ref{sec:strategy_identification}. Games were grouped into \emph{good} and \emph{bad} outcomes based on score at the end of each game. Games with scores above a threshold were classified as good outcomes, whereas games with scores below the threshold were classified as bad outcomes. For each such game, we examined the strategy adopted in the immediately following game to assess how outcome feedback shaped subsequent decisions.

\subsection{Strategy Identification}
\label{sec:strategy_identification}
We recorded complete action trajectories from each game along with the corresponding outcomes (game scores). Each trajectory consists of a sequence of defense and offense decisions made by participants across the test games, resulting in temporally ordered, high-dimensional behavioral data.

For downstream analysis, we reduced each trajectory to a \emph{strategy-level label} based on observed patterns of decision-making. Although this reduction discards fine-grained temporal information, it improves interpretability, facilitates comparisons across participants, and emphasizes higher-level decision patterns, thereby improving the effective signal-to-noise ratio.

\paragraph{Strategy-Level Labeling.}
We assigned each trajectory a strategy-level label based on both the \emph{distribution} and \emph{timing} of aircraft deployment across defense and offense regions. At the distribution level, trajectories were categorized along a continuum from \textcolor{risk}{\emph{risk-seeking}} to \textcolor{balanced}{\emph{balanced}} to \textcolor{loss}{\emph{loss-averse}}, reflecting the trade-off between allocating aircraft offensively versus retaining aircraft to defend the ally carrier. Risk-seeking strategies prioritize offense, whereas loss-averse strategies emphasize defense.

Trajectories classified as \emph{balanced} at the distribution level were further subcategorized based on deployment timing. We characterized defensive deployment timing along two dimensions: initiation timing and deployment aggregation. Initiation timing is categorized as \emph{immediate} if aircraft are deployed at the time of adversary arrival and \emph{delayed} otherwise. Deployment aggregation is categorized as \emph{gradual} if aircraft are deployed across multiple decision steps and \emph{batch} if multiple aircraft are deployed simultaneously. Defense timing was divided into three ordered levels, ranging from \textcolor{balanced}{\emph{delayed–gradual}} deployment (more risk-seeking), to \textcolor{balanced}{\emph{mixed timing}} (delayed–batch or immediate–gradual), to \textcolor{balanced}{\emph{immediate–batch}} deployment (more loss-averse). Offense timing was categorized as either \textcolor{balanced}{\emph{early}} (risk-seeking) or \textcolor{balanced}{\emph{late}} (loss-averse) deployment. These timing-based labels were defined only for trajectories in the balanced distribution group.

In addition, we defined an \textcolor{erratic}{\emph{erratic}} decision-making category to capture trajectories with highly irregular deployment timing, which occurred exclusively within the balanced group. Erratic decision-making describes cases in which aircraft were deployed even when no adversary aircraft were present in the region, leading to unnecessary defensive allocation. Although this behavior indicates a strong defensive tendency, it does not align with the structured risk-to–loss-averse spectrum. We therefore treat it as a qualitatively distinct strategy and label it as $S\text{-}5$. The final strategy labels consisted of six categories, capturing overall allocation patterns and timing differences within balanced strategies:
\begin{itemize}
    \item $S\text{-}0$: risk-seeking;
    \item ($S\text{-}1$)--($S\text{-}3$): balanced strategies, ordered from more risk-seeking to more loss-averse;
    \item $S\text{-}4$: loss-averse;
    \item $S\text{-}5$: erratic decision-making.
\end{itemize}

\begin{figure}[h]
  \centering
  \includegraphics[width=\linewidth]{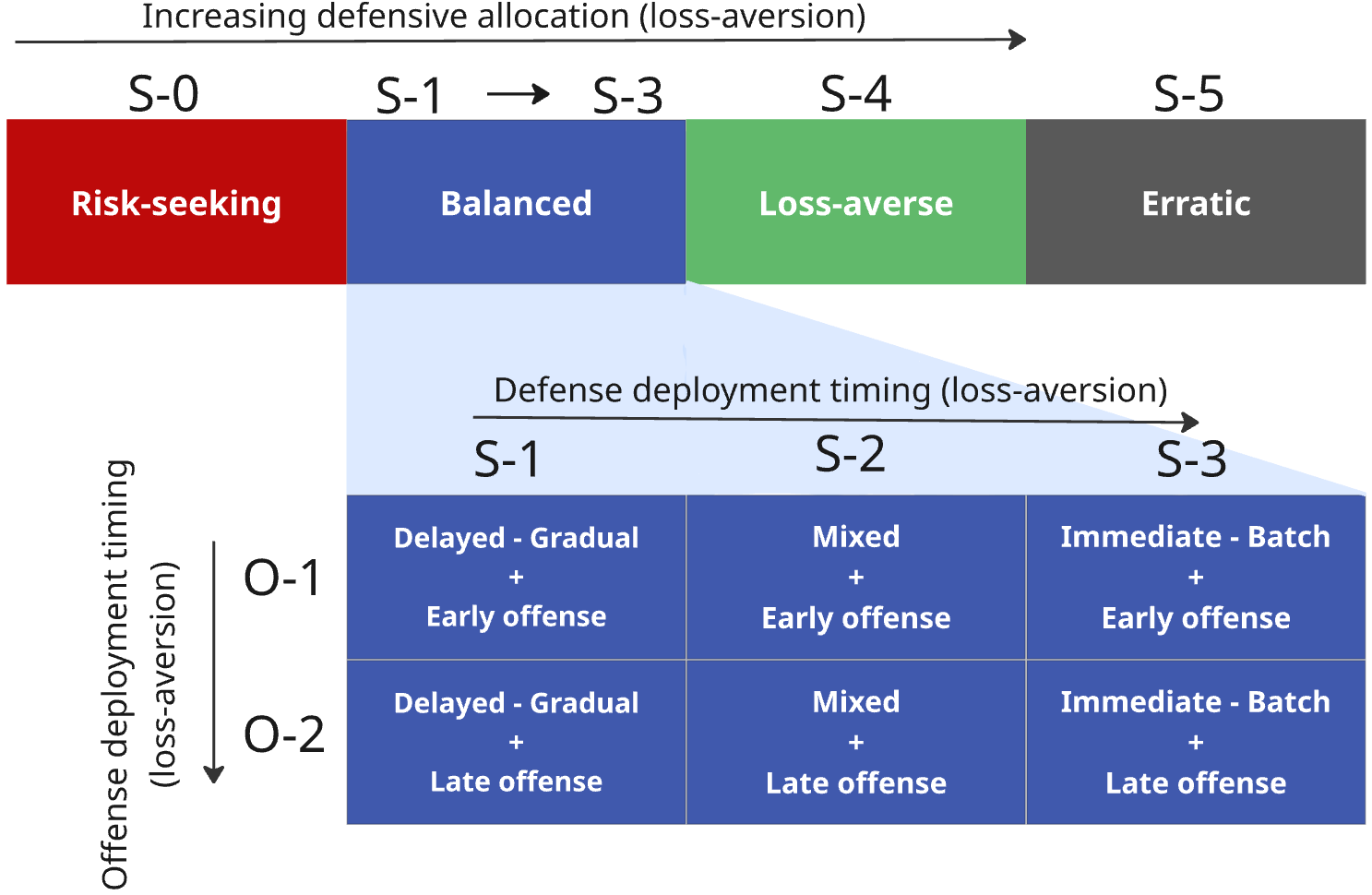}
    \caption{Strategy labeling scheme showing overall strategy labels ($S$-0 to $S$-5). Strategies are arranged from risk-seeking to increasingly loss-averse along the horizontal axis, with an additional erratic category ($S$-5) representing qualitatively distinct defensive behavior. Within the balanced strategies, offense deployment timing labels ($O$-1 and $O$-2) are shown along the vertical axis. Colored regions indicate risk-seeking (red), balanced (blue), loss-averse (green), and erratic (brown) strategy categories.}
    \Description{Diagram showing the strategy labeling scheme used in the study. The top section presents overall strategy categories arranged from left to right according to increasing defensive allocation (loss aversion): S-0 (risk-seeking), S-1 to S-3 (balanced), S-4 (loss-averse), and S-5 (erratic). The balanced category is expanded in the lower section to illustrate timing-based subcategories. Across columns (S-1 to S-3), defensive deployment timing ranges from delayed–gradual to immediate–batch. Along the vertical axis, offense deployment timing varies from early (O-1) to late (O-2). Each cell represents a specific combination of defensive and offensive deployment timing.}

  \label{fig:strategy_labels}
\end{figure}

For trajectories classified as balanced, offense deployment timing was labeled as:
\begin{itemize}
    \item $O\text{-}1$: early (risk-seeking) deployment
    \item $O\text{-}2$: late (loss-averse) deployment
\end{itemize}
A flowchart and summary table of the labeling scheme are provided in Figure~\ref{fig:strategy_labels}.

\paragraph{AI Strategy and Participant Decision Making.}
The AI agents provide recommendations that are empirically sub-optimal with respect to observed game scores and do not maximize reward across all game difficulty levels. Specifically, the DefenseAI follows a strategy corresponding to overall strategy label~$S$-3, while the OffenseAI follows an early offense timing strategy ($O$-1). Participants can achieve higher rewards when they correctly identify when to follow or override AI recommendations, rather than assuming that the AIs are suggesting optimal actions.

Due to the underlying game dynamics, performance is especially sensitive to the timing of defensive deployment. Deploying aircraft to the defense region late and gradually—corresponding to overall strategy label~$S$-1—was empirically associated with higher scores. In contrast, the timing of offensive deployment has a comparatively smaller effect on outcomes. Accordingly, our strategy labels capture meaningful variation in how participants allocate resources across regions in response to sub-optimal AI recommendations.

\begin{figure*}
  \centering
  \includegraphics[width=0.85\linewidth]{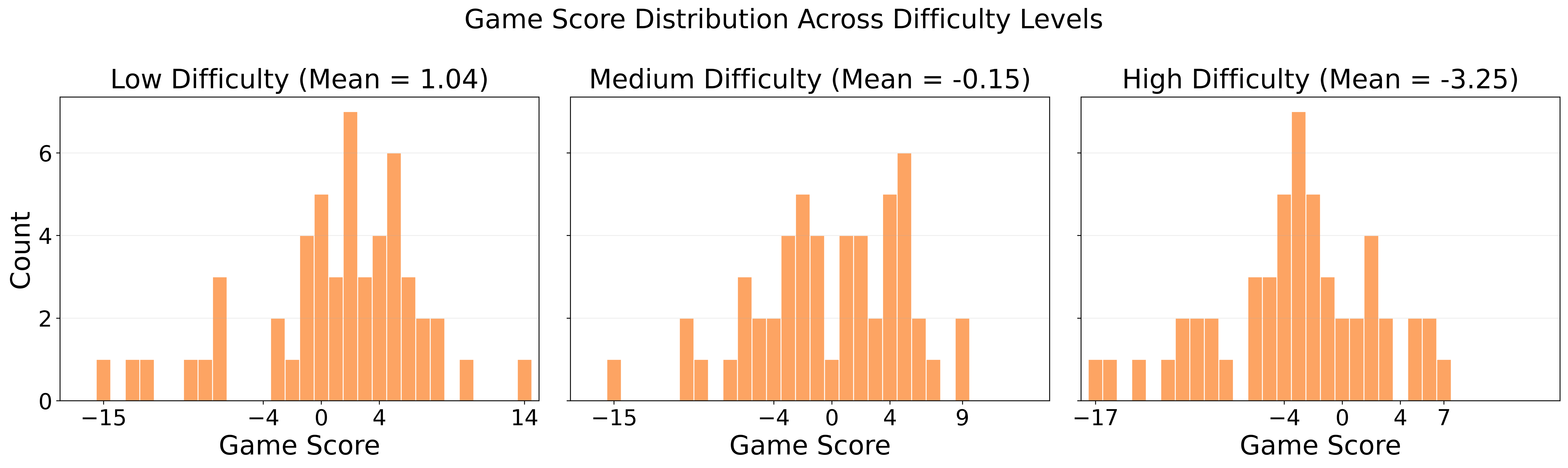}
    \caption{Histogram of game scores by difficulty level across all games (26 participants, 6 games each). The score distribution for low difficulty is shifted toward positive values, whereas high difficulty is shifted toward negative values.}
    \Description{Histogram of game scores by difficulty level across all games (26 participants, 6 games each). The score distribution for low difficulty is shifted toward positive values, whereas high difficulty is shifted toward negative values.}
  \label{fig:game_score_distri}
\end{figure*}

\section{Results}

Each participant’s gameplay trajectory was labeled using the strategy definitions introduced in the previous section. 
As described in Section~\ref{sec:data_collection}, games were categorized as \emph{good} or \emph{bad} outcomes based on the game end scores. 
The distribution of game scores is shown in Figure~\ref{fig:game_score_distri}.

Sections~\ref{sec:results_transitions} and~\ref{sec:results_magnitude} examine changes in participants’ overall strategy classifications, characterizing how strategies shifted following good versus bad outcomes and how the magnitude and direction of these shifts depended on outcome severity. Section~\ref{sec:results_timing} extends this analysis by examining another decision dimension, namely offense deployment timing, to identify outcome-dependent adjustments that are not captured by changes in overall strategy labels.

\begin{table*}[t]
\centering
\scriptsize
\setlength{\tabcolsep}{4pt}
\caption{Poisson regression of strategy transition counts by prior strategy and outcome.}
\label{tab:poisson_interaction}
\begin{tabular}{lcccc}
\toprule
Predictor & Est. & SE & 95\% CI & $p$ \\
\midrule
Intercept (S-4 loss-averse, bad outcome) & 0.2877 & 0.500 & [-0.69, 1.27] & .565 \\

S-0 (risk-seeking)& 1.0986 & 0.548 & [0.03, 2.17] & \textbf{.045*} \\
S-1 (balanced allocation \& delayed-gradual timing) & -0.6931 & 0.339 & [-1.36, -0.03] & \textbf{.041*} \\
S-2 (balanced allocation \& mixed timing) & 0.1178 & 0.645 & [-1.15, 1.38] & .855 \\
S-3 (balanced allocation \& immediate-batch timing) & 1.3610 & 0.537 & [0.31, 2.41] & \textbf{.011*} \\
S-5 (erratic)& 0.2719 & 0.627 & [-0.96, 1.50] & .664 \\

Outcome (Good) & 1.0986 & 0.612 & [-0.10, 2.30] & .073 \\

\midrule
\textit{Strategy $\times$ Outcome (Good)} \\

S-0 (risk-seeking)$\times$ Good Outcome & -1.7918 & 0.742 & [-3.25, -0.34] & \textbf{.016*} \\
S-1 (balanced allocation \& delayed-gradual timing) $\times$ Good Outcome & -0.6931 & 0.339 & [-1.36, -0.03] & \textbf{.041*} \\
S-2 (balanced allocation \& mixed timing) $\times$ Good Outcome & -0.0690 & 0.768 & [-1.57, 1.44] & .928 \\
S-3 (balanced allocation \& immediate-batch timing) $\times$ Good Outcome & -1.1891 & 0.683 & [-2.53, 0.15] & .082 \\
S-5 (erratic)$\times$ Good Outcome & --- & --- & --- & --- \\

\bottomrule
\end{tabular}

\vspace{0.2em}
\footnotesize
\textit{Note.} Reference category: Strategy S-4 (loss-averse strategy) after bad outcome. $^{*}p<.05$.
\end{table*}

\begin{figure}[h]
  \centering
  \includegraphics[width=\linewidth]{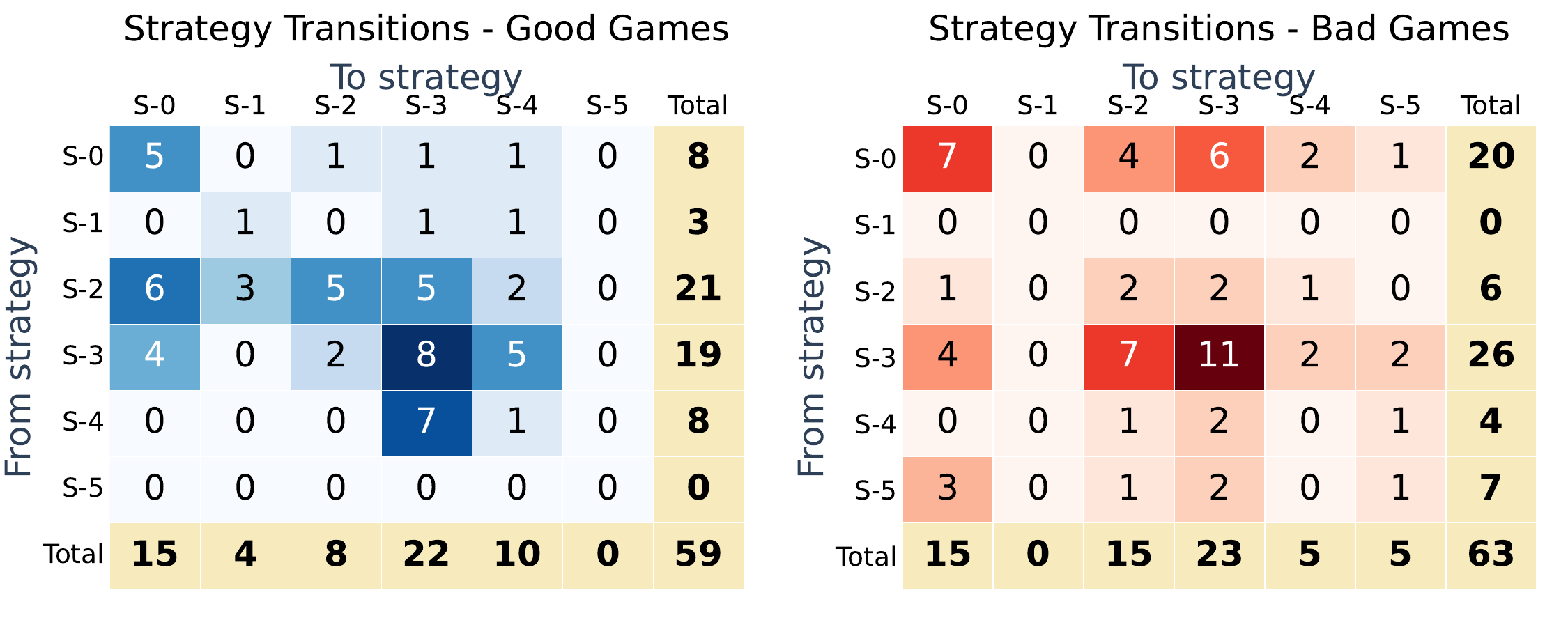}
  \caption{Strategy transition matrices following good and bad games. Each matrix shows how participants transitioned between strategy labels (S-0 to S-5) from one game to the next, conditioned on whether the previous game outcome was good or bad. Cell values indicate the number of observed transitions.}
  \Description{Two heatmaps showing transitions between defense strategy labels across consecutive games. The left heatmap corresponds to transitions following good games, and the right heatmap corresponds to transitions following bad games. Rows represent the strategy used in the previous game, columns represent the strategy used in the subsequent game, and numbers within cells indicate transition counts.}
  \label{fig:heatmap_analysis1}
\end{figure}

\subsection{Outcome-dependent strategy transitions}
\label{sec:results_transitions}

We asked whether participants changed their strategies differently after a good outcome versus a bad outcome. To do so, we grouped games by outcome using score thresholds that matched what participants saw at the end of each game. Scores are displayed as non-negative values in the end-of-game dialog (negative scores are shown as $0$). We therefore defined \emph{good} games as those with a displayed score of at least $1$, and \emph{bad} games as those with a displayed score of $0$. This allowed us to compare behavior following a clearly positive outcome versus an outcome that participants experienced as zero, while avoiding overly strict thresholds that would lead to sparse transition counts.

Figure~\ref{fig:heatmap_analysis1} shows the strategy transition matrices conditioned on whether the previous game outcome was good or bad. After bad outcomes, large shifts were most pronounced among participants using the risk-seeking strategy S-0 and the intermediate strategy S-3, who frequently moved to different strategies in the subsequent game. In contrast, participants using the loss-averse strategy S-4 showed relatively little movement following bad outcomes, with most remaining in the same strategy. After good outcomes, transitions were generally smaller for the risk-seeking strategy S-0 and the intermediate strategy S-3, indicating greater strategy stability following success. For the loss-averse strategy S-4, behavior remained largely stable across both good and bad outcome conditions, with only modest shifts observed after good outcomes, primarily toward more risk-seeking strategies.

To assess whether outcomes influenced how often participants changed strategies, we fit Poisson regression models to the transition counts and compared models with and without an interaction between outcome and starting strategy. A likelihood ratio test revealed a significant interaction between outcome and strategy ($\chi^2(3)=11.848$, $p=0.008$), showing that for some strategies, outcomes led to much more switching, whereas for others outcomes had little effect.

For ease of comparison, we used the loss-averse strategy S-4 as the reference category. This strategy exhibited nonzero transitions in both outcome conditions, while the most extreme strategy (S-5) was absent following good outcomes and therefore could not be reliably estimated. Using S-4 as the baseline allowed us to compare how switching behavior from more risk-seeking and intermediate strategies differed relative to a reference point which is loss averse.

The regression results (Table~\ref{tab:poisson_interaction}) show that after bad outcomes, participants starting in the risk-seeking strategy S-0 switched significantly more often compared to those starting in the loss-averse strategy S-4 ($\beta = 1.10$, $p = .045$). Similarly, participants in the intermediate strategy S-3 exhibited substantially higher transition rates compared to S-4 following bad outcomes ($\beta = 1.36$, $p = .011$).

After good outcomes, these differences were reduced (bottom half of Table~\ref{tab:poisson_interaction}). In particular, the elevated switching from risk-seeking S-0 relative to loss-averse S-4 decreased significantly ($\beta = -1.79$, $p = .016$), indicating less corrective movement following success. For the intermediate strategy S-3, the reduction in switching after good outcomes was smaller and not statistically significant ($\beta = -1.19$, $p = .082$).

For the loss-averse strategy S-4, switching showed a tendency to increase after good outcomes, primarily toward more risk-seeking strategies, but this effect did not reach statistical significance ($\beta = 1.10$, $p = .073$), suggesting that behavior in this strategy remained relatively stable across outcome conditions.

Together, these findings suggest that negative outcomes trigger strong strategy adjustments, particularly away from risk-seeking and intermediate strategies, whereas positive outcomes are associated with greater stability and more modest shifts in behavior.

\begin{figure}[h]
  \centering
  \includegraphics[width=\linewidth]{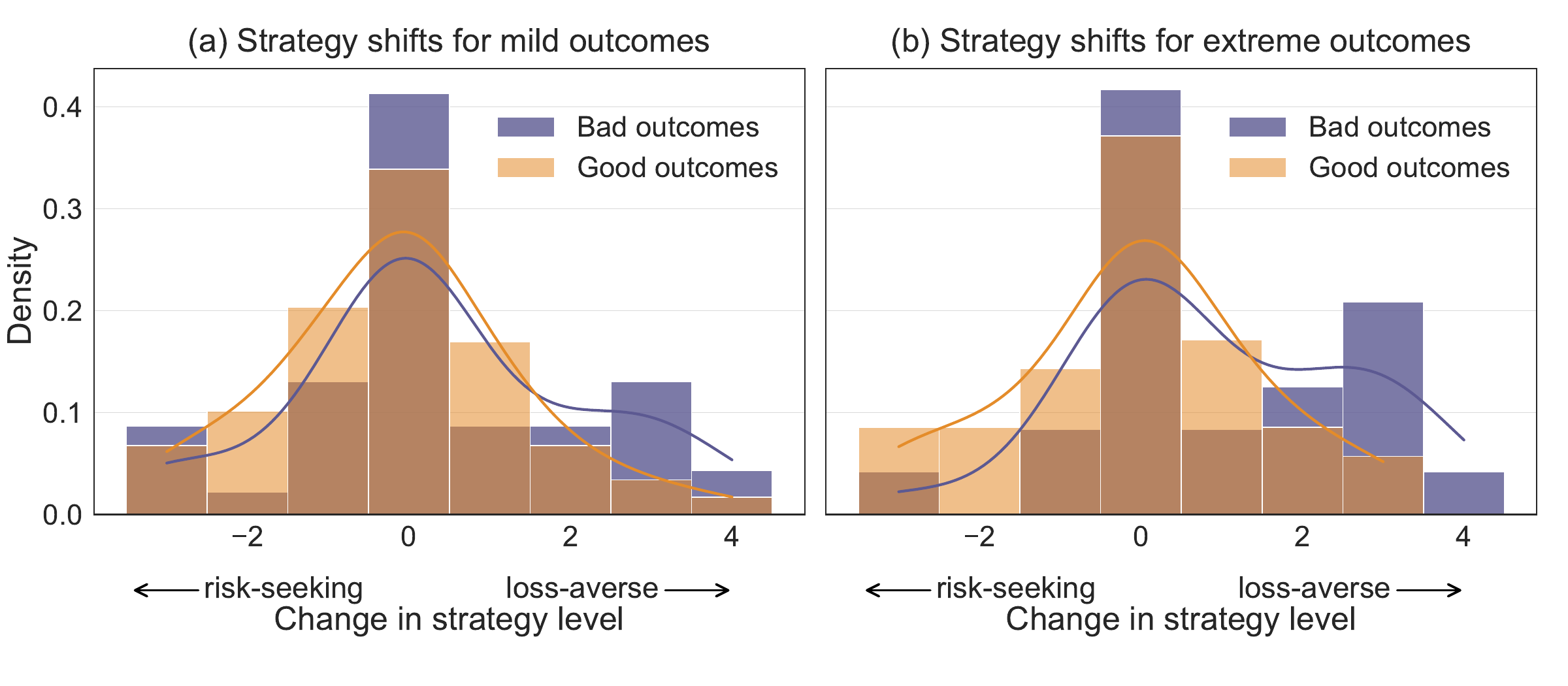}
  \caption{
    Density distributions of signed changes in strategy labels between consecutive games.
    Negative values indicate shifts toward more risk-seeking strategies, whereas positive values indicate shifts toward greater loss aversion.
    (a) Moderate outcomes (good: score $\geq 1$, bad: score $\leq -1$).
    (b) Extreme outcomes (good: score $\geq 4$, bad: score $\leq -4$).
    }
   \Description{Two side-by-side density histograms showing changes in strategy between consecutive games. 
    The horizontal axis shows signed changes in strategy labels, and the vertical axis shows density. 
    Each histogram overlays distributions for good outcomes and bad outcomes. 
    The left histogram corresponds to moderate outcomes, and the right histogram corresponds to extreme outcomes.}
  \label{fig:analysis_2_density}
\end{figure}

\subsection{Direction and magnitude of outcome-driven strategy shifts}
\label{sec:results_magnitude}
To more directly examine how outcome severity influences subsequent behavior, we focus on games with extreme outcomes, using score thresholds of +4 and –4. This choice is motivated by how outcome feedback is perceived during gameplay. At the end of each game, scores are presented in a dialog box as truncated values, which makes small differences in health (e.g., –1 or –2) less perceptible to participants. In contrast, when outcomes are strongly negative or positive, participants can often anticipate the result before the game ends by monitoring the evolving state of their aircraft and the adversary forces. As a result, large losses and gains tend to become salient during play, rather than only when the final score is displayed.

We additionally selected thresholds of +4 and –4 because such outcomes are more likely to carry a consistent subjective interpretation across participants. Smaller score differences (e.g., –1 or +1) may be evaluated differently depending on individual expectations, whereas outcomes of +4 or –4 are more likely to be uniformly perceived as substantial gains or losses. Using these more extreme thresholds therefore provides a cleaner and more interpretable basis for isolating behavioral responses to clearly positive versus clearly negative outcomes.

Because strategy labels $S$-0 through $S$-4 form an ordered spectrum from risk-seeking to loss-averse behavior, whereas $S$-5 represents erratic decision-making rather than a position on this spectrum, we excluded $S$-5 from this analysis. This allowed us to quantify both the magnitude and direction of strategy shifts along a single interpretable dimension. For each pair of consecutive games, we computed the strategy shift as the signed difference between the strategy label in the later game and the strategy label in the preceding game. For example, a change from $S$-4 to $S$-1 corresponds to a value of $-3$, indicating a shift toward more risk-seeking behavior, whereas positive values indicate movement toward greater loss aversion.

Figure~\ref{fig:analysis_2_density} show the density distributions of strategy shifts following positive and negative outcomes. In both figures, the horizontal axis represents the signed difference in strategy labels, with negative values indicating shifts toward more risk-seeking strategies and positive values indicating shifts toward greater loss aversion. Figure~\ref{fig:analysis_2_density}(a) uses a looser definition of outcomes (good: scores $\geq 1$, bad: scores $\leq -1$), whereas Figure~\ref{fig:analysis_2_density}(b) focuses on more extreme gains (scores $\geq 4$) and losses (scores $\leq -4$).

As shown in Figure~\ref{fig:analysis_2_density}(b), strongly negative outcomes are associated with a rightward shift in the distribution of strategy changes relative to Figure~\ref{fig:analysis_2_density}(a), indicating a systematic movement toward more loss-averse strategies following large losses. This pattern reflects loss-averse behavior, consistent with prior work \cite{10.1145/3410404.3414250}. In contrast, the distribution of strategy shifts following positive outcomes remains largely unchanged across threshold definitions.

To assess whether strategy shifts differed following extreme losses versus gains, we compared the distributions of strategy changes using a Mann–Whitney U test. When using mild outcome thresholds (scores $\geq 1$ vs.\ $\leq -1$), the difference between conditions was not significant ($U = 1582$, $p = 0.135$, $r = -0.17$), indicating no reliable shift in strategy following ambiguous outcomes.

In contrast, when restricting the analysis to extreme outcomes (scores $\geq 4$ vs.\ $\leq -4$), the distributions differed significantly ($U = 547$, $p = 0.043$, $r = -0.30$). This result indicates that participants were more likely to shift toward more loss-averse strategies following large negative outcomes. Together, these findings suggest that outcome severity plays a critical role in driving systematic changes in strategy.

In this analysis, we consider only the overall strategy labels ($S$-0 through $S$-4), which capture changes along the risk-seeking to loss-averse spectrum but do not distinguish between offense deployment timing strategies ($O$-1 and $O$-2). Accordingly, the prominent peak at zero in the strategy-shift distributions indicates that many participants did not change their overall strategy label across consecutive games. However, offense timing constitutes a separate behavioral dimension along which risk-seeking or loss-averse responses may also be expressed. As a result, participants may exhibit meaningful shifts in offense deployment timing—even when no change is observed at the level of overall strategy labels. We examine such outcome-dependent adjustments in offense deployment timing in the following analysis.

\begin{figure}[h]
  \centering
  \includegraphics[width=0.85\linewidth]{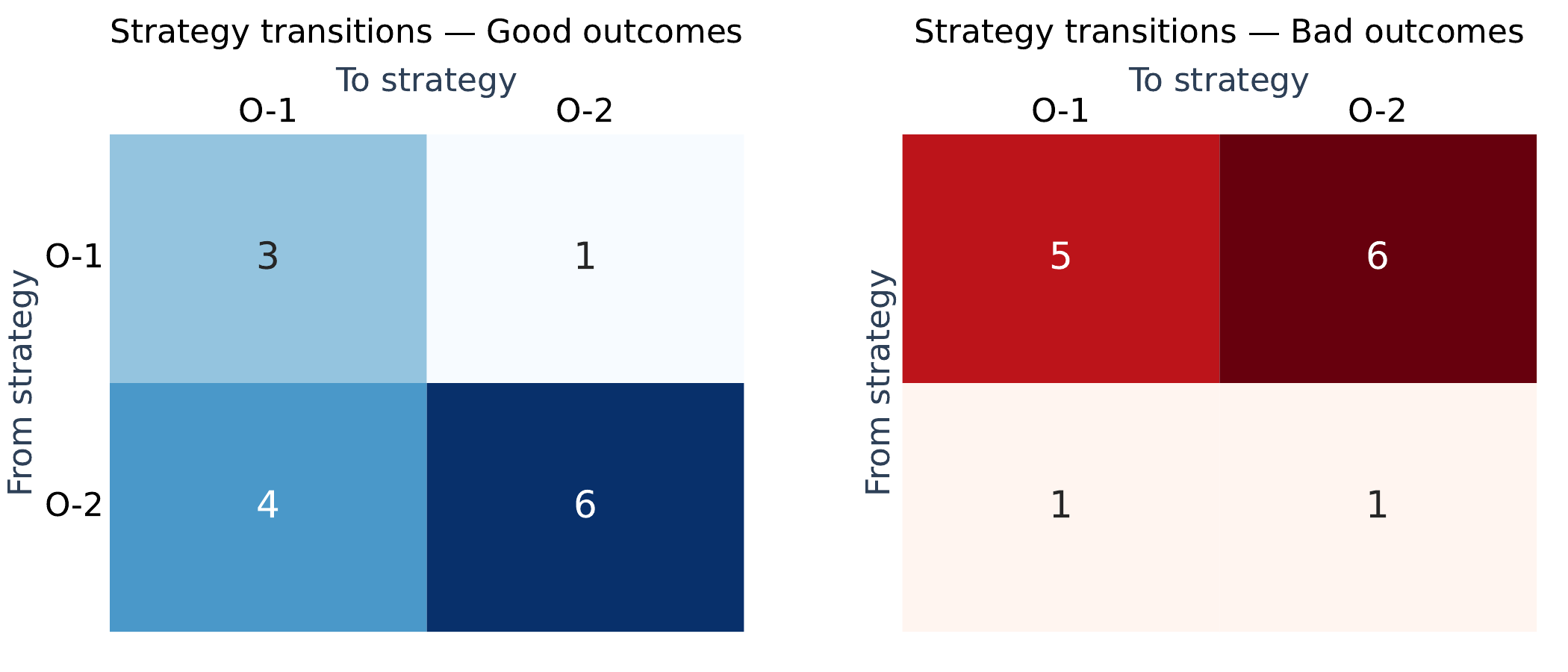}
  \caption{Offense strategy transitions when overall strategy remains unchanged, showing frequent shifts from early ($O$-1) to late ($O$-2) offense deployment following bad outcomes.}
  \Description{Heatmaps showing how participants transition between offense strategies from one game to the next for good and bad outcomes, when their overall strategy does not change.}

  \label{fig:analysis_3_filtered}
\end{figure}

\begin{table}[t]
\centering
\scriptsize
\setlength{\tabcolsep}{4pt}
\caption{Poisson regression of offense timing transitions by prior offense strategy and outcome.}
\label{tab:poisson_offense}
\begin{tabular}{lcccc}
\toprule
Predictor & Est. & SE & 95\% CI & $p$ \\
\midrule
Intercept (O-1 early offense, bad outcome) & 1.7047 & 0.302 & [1.11, 2.30] & \textbf{<.001} \\

 O-2 (late offense) & -1.7047 & 0.769 & [-3.21, -0.20] & \textbf{.027*} \\
Outcome (Good) & -1.0116 & 0.584 & [-2.16, 0.13] & .083 \\

\midrule
\textit{Strategy $\times$ Outcome (Good)} \\
O-2 (late offense) $\times$ Good Outcome& 2.6210 & 0.970 & [0.72, 4.52] & \textbf{.007*} \\

\bottomrule
\end{tabular}

\vspace{0.2em}
\footnotesize
\textit{Note.} Reference category: O-1 (early offense) after bad outcome. $^{*}p<.05$.
\end{table}
\subsection{Outcome-dependent offense timing adjustments}
\label{sec:results_timing}

The previous analyses showed that many participants did not change their overall defensive strategy across consecutive games, even after extreme negative outcomes. However, participants could also adjust their offense deployment timing. Here, we examine whether offense timing transitions differed following good versus bad outcomes.

Figure~\ref{fig:analysis_3_filtered} illustrates offense timing transitions conditioned on prior outcome for games in which participants maintained the same overall defensive strategy. After bad outcomes, participants frequently shifted from early offense deployment ($O$-1) to later deployment ($O$-2), whereas after good outcomes they were more likely to retain their existing offense timing strategy. Early offense can be though of as risk-seeking and late offense is more like loss averse.

To assess whether offense timing adjustments depended on outcome, we fit Poisson regression models to transition counts and compared models with and without an interaction between offense strategy and outcome. A likelihood ratio test revealed a significant interaction ($\chi^2(1)=9.18$, $p=0.002$), indicating that outcome influenced how often participants changed offense timing.

Regression results (Table~\ref{tab:poisson_offense}) show that participants using early offense deployment switched strategies significantly more often after bad outcomes than those using late deployment ($\beta=-1.70$, $p=.027$). After good outcomes, participants were more likely to retain early offense timing, reflecting greater stability following success. This shift is also confirmed by a significant interaction between offense timing and outcome ($\beta=2.62$, $p=.007$)

We conducted a complementary analysis that also included cases in which participants changed their defensive strategy while remaining within the balanced range. Using the same Poisson modeling approach, we again observed a significant interaction between outcome and offense timing transitions (likelihood ratio test: $\chi^2(1)=8.01$, $p=0.005$; Table~\ref{tab:poisson_offense_robust}). 

Consistent with the primary analysis, participants were significantly more likely to change offense timing following bad outcomes when using early offense deployment compared to late deployment ($\beta=-0.98$, $p=.040$). The significant interaction term ($\beta=1.73$, $p=.007$) similarly indicates that the strong tendency to adjust offense timing after bad outcomes was substantially weaker following good outcomes.

Because performance feedback was observed only after a sequence of interdependent actions involving both DefenseAI and OffenseAI, participants appeared to struggle with temporal credit assignment. Rather than adjusting defense-related choices that would have more directly improved performance, they systematically modified offense deployment timing following failure. One plausible explanation is that participants intuitively associated losses with insufficient defense and therefore delayed offense deployment, keeping aircraft in reserve for later stages, without altering their defensive strategy and treating offense timing as the most immediate lever for correction. This indicates biased error attribution under delayed feedback in multi-step human–AI decision-making.

\begin{figure}[h]
  \centering
  \includegraphics[width=0.85\linewidth]{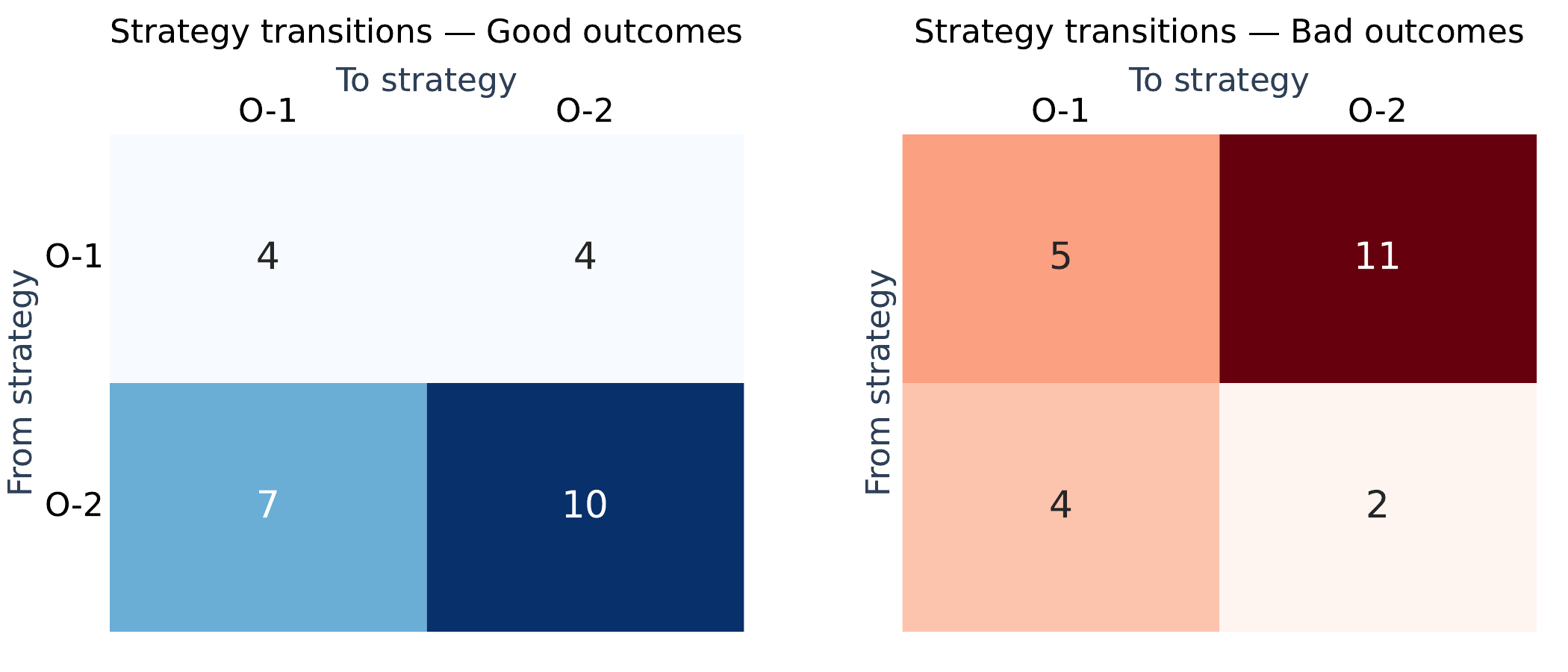}
  \caption{Offense strategy transitions when defensive strategies vary within the balanced range ($S$-1 to $S$-3), showing increased shifts from early ($O$-1) to late ($O$-2) offense deployment following bad outcomes.}
  \Description{Heatmaps showing transitions between offense strategies across consecutive games for good and bad outcomes, including cases where defensive strategies change within the balanced range ($S$-1 to $S$-3).}
  \label{fig:analysis_3_unfiltered}
\end{figure}

\begin{table}[t]
\centering
\scriptsize
\setlength{\tabcolsep}{4pt}
\caption{ Poisson regression of offense timing transitions when defensive strategies vary within the balanced range ($S$-1 to $S$-3).}
\label{tab:poisson_offense_robust}
\begin{tabular}{lcccc}
\toprule
Predictor & Est. & SE & 95\% CI & $p$ \\
\midrule
Intercept (O-1 Early offense, bad) & 2.0794 & 0.250 & [1.59, 2.57] & \textbf{<.001} \\
O-2 (late offense) & -0.9808 & 0.479 & [-1.92, -0.04] & \textbf{.040*} \\
Outcome (Good) & -0.6931 & 0.433 & [-1.54, 0.16] & .109 \\
\midrule
\textit{Strategy $\times$ Outcome (Good)} \\
O-2 (late offense) $\times$ Good & 1.7346 & 0.643 & [0.48, 2.99] & \textbf{.007*} \\
\bottomrule
\end{tabular}

\vspace{0.2em}
\footnotesize
\textit{Note.} Reference category: O-1 (early offense) after bad outcome. $^{*}p<.05$.
\end{table}

\section{Discussion}

This work examined how delayed outcomes shape learning and responsibility attribution in a multi-agent human–AI decision-making task. Participants showed outcome-dependent adjustments, with substantially behavioral changes following negative outcomes than after positive outcomes and exhibited loss-averse behavior consistent with that observed in the game setup in \cite{10.1145/3410404.3414250}.

However, corrective behavior was not aligned with the decisions most responsible for performance. In the defense strategy space, participants frequently failed to move toward strategies that would have directly improved outcomes. In particular, when balanced strategies performed poorly, participants tended to remain near intermediate strategies rather than shifting toward more risk-seeking strategies that would have increased success. This pattern reflects a temporal credit assignment problem: delayed feedback made it difficult for participants to identify which earlier decisions caused failure, leading to weak or incomplete corrective learning.

At the same time, participants systematically adjusted offense deployment timing following negative outcomes, even when their defensive strategy remained unchanged. Shifts from early to late offense deployment were common after failure but much less frequent after success. Although intuitive as a protective response, these timing adjustments were often weakly related to the underlying causes of poor performance. Rather than correcting defense-related decisions that drove outcomes, participants redirected effort toward offense timing.

Together, these patterns reveal a consistent form of biased error attribution under delayed feedback. Participants struggled to identify which components of the decision process were causally responsible for outcomes and instead applied corrections either incompletely (remaining near ineffective defense strategies) or to less relevant decision dimensions (offense timing). Therefore, the presence of cascaded decision steps along with multiple AI agents complicated this attribution process, increasing the likelihood of misdirected learning.

This work bridges cognitive bias research with sequential decision-making under delayed feedback, offering empirical evidence of how loss aversion and attribution bias interact in multi-agent human–AI contexts. More broadly, our results highlight the need to move beyond single-step human–AI interaction models and toward understanding bias dynamics in complex, multi-stage decision environments.

From a design perspective, multi-step human–AI systems with delayed feedback may benefit from interfaces and feedback mechanisms that explicitly highlight the contribution of individual decisions. Such support could help users develop more accurate mental models of system dynamics and AI behavior, thereby promoting more effective learning.

Moreover, because individuals differ in the types and strengths of cognitive biases they exhibit, bias-mitigation mechanisms should be adaptive. Systems that adjust feedback and guidance based on users’ observed behavior may help reduce biased error attribution and support more effective strategy adjustment. Evaluating such personalized mitigation approaches in human–AI decision-making tasks with delayed outcomes and multiple autonomous agents represents an important direction for improving learning and performance in complex decision environments. 

\section{Limitations and Future Work}
One of the limitations of this work is that behavioral inferences were derived from objective gameplay trajectories without analyzing participants’ mental models, although such subjective data were collected but not examined in the present work. Mental model analysis is important for understanding why particular actions were taken in certain situations \cite{andrews2023shared} and would allow us to compare participants’ explanations with their observed behavior, thereby cross-validating our findings for these scenarios. 

A second limitation relates to the definition of the optimal strategy and the ground truth used for claiming the bias error attribution. The optimal deployment strategy used in our analysis is empirical, in the sense that it was identified based on the strategy that performed best across the game scenarios presented to participants, rather than being derived from a formal proof of optimality.

A further limitation arises from how the reference sequence of actions used for analyzing error attribution was defined. In our study, the AI agent was trained without a human in the loop, and the sequence of actions suggested by the trained policy in this setting was used as a reference. This sequence does not represent a formally optimal strategy, but rather the behavior learned by the AI during training. Due to differences between the training environment and the software environment in which the agent was deployed, the AI’s suggested actions were not always appropriate, and achieving the best outcome required participants to sometimes follow the AI’s suggestions and at other times override them.

To analyze bias in error attribution, participant decisions were evaluated relative to this reference sequence of AI-suggested actions, and deviations from the expected pattern of following or correcting the AI were treated as attribution errors. However, once a participant deviates from the reference sequence, the game transitions to a different state, after which the AI produces a different set of action suggestions. Consequently, not all participants encountered the same sequence of AI suggestions that defines the reference. This limitation is important when interpreting our claims about bias in error attribution, as these claims are defined relative to the reference sequence obtained from AI-only runs, even though some participants were not exposed to the same decision sequence during gameplay.

Future work should examine longer interaction periods \cite{martin2004actrbeer} and incorporate mental model assessment to determine whether extended experience reduces biased attribution and improves alignment with optimal decision strategies. Future work should also investigate how AI systems behave when humans do not follow the AI’s suggested actions, which may place the AI in environment states outside the training state distribution (out-of-distribution states), and how such situations influence subsequent human decision-making. Such out-of-distribution behavior is common in real-world AI and robotics systems, where models may produce confident responses in unfamiliar states, posing potential safety risks.

\section{Acknowledgements}
This work was supported by the Office of Naval Research Command Decision Making Program under Contract N00014‐24‐S‐B001. The results do not reflect the official position of this agency.

\bibliographystyle{ACM-Reference-Format}
\bibliography{sample-base}

\clearpage
\appendix
\small
\section{Questionnaires}
\label{app:questionnaires}
All questionnaires administered during the study are included below for completeness.

\subsection{Group 1 — Pre-Training Questionnaire}

In this phase, participants answered demographic questions as well as questions about their experience with AI and gaming.

\paragraph{Demographics}

\begin{itemize}
\item What is your age?
\begin{itemize}
\item 18--24
\item 25--39
\item 40--65
\end{itemize}

\item What is your gender?
\begin{itemize}
\item Male
\item Female
\item Non-binary
\item Prefer not to say
\end{itemize}
\end{itemize}

\paragraph{Experience with AI}

\begin{itemize}
\item Have you taken any undergraduate-level course in machine learning?
\begin{itemize}
\item Yes
\item No
\end{itemize}

\item Have you taken any graduate-level course in machine learning?
\begin{itemize}
\item Yes
\item No
\end{itemize}

\item Do you have experience designing AI models either through research or industry experience?
\begin{itemize}
\item Yes
\item No
\end{itemize}
\end{itemize}

\paragraph{Gaming Experience}

\begin{itemize}
\item How often do you play video games on a console, PC, smartphone, or tablet?
\begin{itemize}
\item Almost never
\item A few hours per month
\item A few hours per week
\item More than 5 hours per week
\end{itemize}

\item Which of the following game genres do you (or did you) play often? (Select all that apply)
\begin{itemize}
\item Real-time strategy (e.g., DOTA, League of Legends, Starcraft)
\item Turn-based strategy (e.g., Civilization)
\item Other
\end{itemize}
\end{itemize}

\paragraph{AI's Confidence}

\begin{itemize}
\item Based on prior experience with AI, what level of model confidence is sufficient for you to rely on the AI's output?

\begin{itemize}
\item At least 50\%
\item At least 70\%
\item At least 90\%
\item 100\%
\item Doesn't matter, I will always evaluate the decision irrespective of the confidence level
\end{itemize}

\end{itemize}

\subsection{Group 1 — Post-Training Questionnaire}

In this phase, participants were asked about their confidence in playing the game without external assistance.

\begin{itemize}

\item How would you describe your approach to risk in this game?

\begin{itemize}
\item I take balanced risks to aim for a reasonable reward.  
(Balanced risk means making decisions that involve some risk, but with a strong chance of achieving a safe minimum target.)
\item I am willing to take higher risks to aim for the maximum reward, even if it means I might end up with a lower amount.
\end{itemize}

\item How confident are you in your ability to win the game based solely on your own decisions, without any external suggestions or assistance?  
Rate on a scale from 1 to 5, where 1 means least confident and 5 means very confident.

\begin{itemize}
\item 1
\item 2
\item 3
\item 4
\item 5
\end{itemize}

\end{itemize}

\subsection{Group 1 — Test Game Questionnaire}

Participants played multiple games, and the following questions were asked after each game.

\begin{itemize}

\item How much do you trust DefenseAI?  
Rate on a scale from 1 to 5, where 1 means lowest trust and 5 means full trust.

\begin{itemize}
\item 1
\item 2
\item 3
\item 4
\item 5
\end{itemize}

\item How much do you trust OffenseAI?  
Rate on a scale from 1 to 5, where 1 means lowest trust and 5 means full trust.

\begin{itemize}
\item 1
\item 2
\item 3
\item 4
\item 5
\end{itemize}

\item At any point during the current game, did DefenseAI provide a completely incorrect suggestion?

\begin{itemize}
\item Yes
\item No
\item I'm not sure
\item I did not pay attention to the AI's suggestion, so I am not sure
\end{itemize}

\item At any point during the current game, did OffenseAI provide a completely incorrect suggestion?

\begin{itemize}
\item Yes
\item No
\item I'm not sure
\item I did not pay attention to the AI's suggestion, so I am not sure
\end{itemize}

\end{itemize}

\paragraph{Mental Model Questionnaire}

The following questions were asked after each game to understand how participants interpreted the AI advisors’ behavior.
\begin{itemize}
\item Was there any moment when you felt that the DefenseAI's suggestion might have made you lose or perform worse if you had followed it exactly?  
Or, if you did follow the DefenseAI's suggestion, what would you do differently next time to perform better?

\begin{itemize}
\item Open-ended response
\end{itemize}

\item Was there any moment when you felt that the OffenseAI's suggestion might have made you lose or perform worse if you had followed it exactly?  
Or, if you did follow the OffenseAI's suggestion, what would you do differently next time to perform better?

\begin{itemize}
\item Open-ended response
\end{itemize}

\end{itemize}

\subsection{Group 1 — Final Questionnaire}

At the end of the experiment, participants answered the following questions.

\begin{itemize}

\item Based on your experience with the Defense AI during all the games, how much would you trust it for decision-making in the future?

Rate on a scale from 1 to 5.

\begin{itemize}
\item 1
\item 2
\item 3
\item 4
\item 5
\end{itemize}

\item Based on your experience with the Offense AI during all the games, how much would you trust it for decision-making in the future?

Rate on a scale from 1 to 5.

\begin{itemize}
\item 1
\item 2
\item 3
\item 4
\item 5
\end{itemize}

\item Please explain your response for the above two questions.

\begin{itemize}
\item Open-ended response
\end{itemize}

\item After playing all the games, do you think you were able to figure out a strategy to win (or perform high) in the game?

\begin{itemize}
\item Yes
\item No
\end{itemize}

\item Please explain your response for the above question.

\begin{itemize}
\item Open-ended response
\end{itemize}

\item At any time during any of the games, did you feel the need to control which targets the aircraft were engaging?

\begin{itemize}
\item Yes
\item No
\end{itemize}

\item Did this lack of control affect your reliance on AIs?

\begin{itemize}
\item Yes
\item No
\end{itemize}

\item Please explain your responses for the above questions.

\begin{itemize}
\item Open-ended response
\end{itemize}

\item Which of the following should the AIs have communicated, in addition to their suggestions? (Select all that apply.)

\begin{itemize}
\item Confidence score for each suggestion
\item Rationale behind each suggestion
\item What each AI considered about its region
\item Whether they are communicating with or aware of other regions
\item The lookahead steps they are considering
\item Other (please specify)
\end{itemize}

If other, please specify:
\begin{itemize}
\item Open-ended response
\end{itemize}

\item How did you find the pace of decision-making (30s per decision)?

\begin{itemize}
\item 30s was too short to make a decision
\item 30s was reasonable for making a decision on average
\item 30s was more than enough to make a decision
\end{itemize}

\item How mentally demanding was it to make decisions during the games?

Rate on a scale from 1 to 10.

\begin{itemize}
\item 1
\item 2
\item 3
\item 4
\item 5
\item 6
\item 7
\item 8
\item 9
\item 10
\end{itemize}

\end{itemize}

\subsection{Group 2 — Pre-Training Questionnaire}

In this phase, participants answered demographic questions as well as questions about their experience with AI and gaming.

\paragraph{Demographics}

\begin{itemize}

\item What is your age?
\begin{itemize}
\item 18--24
\item 25--39
\item 40--65
\end{itemize}

\item What is your gender?
\begin{itemize}
\item Male
\item Female
\item Non-binary
\item Prefer not to say
\end{itemize}

\end{itemize}

\paragraph{Experience with AI}

\begin{itemize}

\item Have you taken any undergraduate-level course in machine learning?
\begin{itemize}
\item Yes
\item No
\end{itemize}

\item Have you taken any graduate-level course in machine learning?
\begin{itemize}
\item Yes
\item No
\end{itemize}

\item Do you have experience designing AI models either through research or industry experience?
\begin{itemize}
\item Yes
\item No
\end{itemize}

\end{itemize}

\paragraph{Gaming Experience}

\begin{itemize}

\item How often do you play video games on a console, PC, smartphone, or tablet?
\begin{itemize}
\item Almost never
\item A few hours per month
\item A few hours per week
\item More than 5 hours per week
\end{itemize}

\item Which of the following game genres do you (or did you) play often? Select all that apply.
\begin{itemize}
\item Real time strategy (e.g., DOTA, League of Legends, Starcraft)
\item Turn-based strategy (e.g., Civilization)
\item Other
\end{itemize}

\end{itemize}

\paragraph{AI's Confidence}

\begin{itemize}
\item Based on prior experience with AI, what level of model confidence is sufficient for you to rely on the AI's output?

\begin{itemize}
\item At least 50\%
\item At least 70\%
\item At least 90\%
\item 100\%
\item Doesn't matter, I will always evaluate the decision irrespective of the confidence level
\end{itemize}

\end{itemize}

\subsection{Group 2 — Post-Training Questionnaire}

In this phase, participants were asked about their confidence in playing the game without AI support.
\begin{itemize}

\item How would you describe your approach to risk in this game?

\begin{itemize}
\item I take balanced risks to aim for a reasonable reward.  
(Balanced risk means making decisions that involve some risk, but with a strong chance of achieving a safe minimum target).

\item I'm willing to take higher risks to aim for the maximum reward (\$25), even if it means I might end up with a lower amount (e.g., \$15).
\end{itemize}

\item How confident are you in your ability to play the game without any external assistance?

Rate on a scale from 1 to 5, where 1 means least confident and 5 means very confident.

\begin{itemize}
\item 1
\item 2
\item 3
\item 4
\item 5
\end{itemize}

\end{itemize}

\subsection{Group 2 — Test Game Questionnaire}

Participants played multiple games, and the following questions were asked after each game.

\begin{itemize}

\item How much do you trust Offense AI?  
Rate on a scale of 1--5, with 1 signifying lowest trust and 5 representing full trust.

\begin{itemize}
\item 1
\item 2
\item 3
\item 4
\item 5
\end{itemize}

\item What level of model confidence is sufficient for you to rely on the DefenseAI's output?  
The confidence level shows how sure the AI is that its suggested action will lead to the best long-term outcome.

\begin{itemize}
\item At least 50\%
\item At least 70\%
\item At least 90\%
\item 100\%
\item Doesn't matter, will always evaluate its decision, irrespective of the confidence level.
\end{itemize}

\item What level of model confidence is sufficient for you to rely on the OffenseAI's output?

\begin{itemize}
\item At least 50\%
\item At least 70\%
\item At least 90\%
\item 100\%
\item Doesn't matter, will always evaluate its decision, irrespective of the confidence level.
\end{itemize}

\item At any point during the current game, did DefenseAI provide a completely incorrect suggestion with high confidence?

\begin{itemize}
\item Yes
\item No
\item I'm not sure
\item I didn't pay attention to the AI's confidence levels that is why I'm not sure
\end{itemize}

\item At any point during the current game, did OffenseAI provide a completely incorrect suggestion with high confidence?

\begin{itemize}
\item Yes
\item No
\item I'm not sure
\item I didn't pay attention to the AI's confidence levels that is why I'm not sure
\end{itemize}

\end{itemize}

\paragraph{Mental Model Questionnaire}

The following questions were asked after each game to understand how participants interpreted the AI advisors’ behavior.
\begin{itemize}
\item Was there any moment when you felt that the DefenseAI's suggestion might have made you lose or perform worse if you had followed it exactly?  
Or, if you did follow the DefenseAI's suggestion, what would you do differently next time to perform better?

\begin{itemize}
\item Open-ended response
\end{itemize}

\item Was there any moment when you felt that the OffenseAI's suggestion might have made you lose or perform worse if you had followed it exactly?  
Or, if you did follow the OffenseAI's suggestion, what would you do differently next time to perform better?

\begin{itemize}
\item Open-ended response
\end{itemize}

\end{itemize}

\subsection{Group 2 — Final Questionnaire}

At the end of the experiment, participants answered the following questions.

\begin{itemize}

\item Overall, how much do you trust DefenseAI?

Rate on a scale from 1 to 5, where 1 means lowest trust and 5 means full trust.

\begin{itemize}
\item 1
\item 2
\item 3
\item 4
\item 5
\end{itemize}

\item Overall, how much do you trust OffenseAI?

Rate on a scale from 1 to 5, where 1 means lowest trust and 5 means full trust.

\begin{itemize}
\item 1
\item 2
\item 3
\item 4
\item 5
\end{itemize}

\item Have you figured out a strategy that helps you perform well in the game?
If yes, please describe it.

\begin{itemize}
\item Open-ended response
\end{itemize}

\item As a commander, you were responsible for high-level planning while low-level control was handled automatically.
Did this limitation affect how much you relied on the AI?

\begin{itemize}
\item Yes
\item No
\item Not sure
\end{itemize}

\item Which of the following should the AIs have communicated, in addition to their suggestions? Select all that apply.

\begin{itemize}
\item Rationale behind each suggestion
\item What each AI considered about its region
\item Whether they are communicating with or aware of other regions
\item The lookahead steps they are considering
\item Other, please specify
\end{itemize}

\item How did you find the pace of decision-making (30s per decision)?

\begin{itemize}
\item 30s was too short to make a decision
\item 30s was reasonable for making a decision on average
\item 30s was more than enough to make a decision
\end{itemize}

\item How mentally demanding was it to make decisions during the games?

\begin{itemize}
\item 1
\item 2
\item 3
\item 4
\item 5
\item 6
\item 7
\item 8
\item 9
\item 10
\end{itemize}

\end{itemize}




\end{document}